\documentclass[12pt]{article}
\usepackage{cmb,epsfig}

\begin{document}

\heading{THE OBSERVABILITY OF SECONDARY DOPPLER PEAKS}

\author{Michael P. Hobson} {Mullard Radio Astronomy Observatory, Cambridge,
UK.} {$\;$}

\begin{abstract}{\baselineskip 0.4cm
By proposing
a statistic for the detection of secondary (Doppler) peaks in the CMBR
power spectrum, the significance level at which such peaks may be
detected are computed for a large range of model CMBR experiments.  In
particular, we investigate experimental design features required to
distinguish between competing cosmological theories, such as cosmic
strings and inflation, by establishing whether or not secondary peaks
are present in the CMBR power spectrum.}
\end{abstract}

\section{Introduction}

Experimental measurement of the Doppler peaks' positions and heights
would fix at least some combinations of cosmological parameters
(e.g. $H_0$, $\Omega_0$ etc.)  which are left free in inflationary
models \cite{jungman}.  Furthermore, as shown by 
\cite{albrecht,magueijo}, the {\em absence} of secondary
Doppler peaks is a robust prediction for cosmic strings, although
this may or may not be the case for textures \cite{crittenden,durrer}.
Therefore it appears
that even determining whether or not there are secondary Doppler peaks
offers an important alternative motivation
for measuring the CMBR power spectrum.

We address this issue by proposing a statistic for detecting secondary
oscillations, and studying how it performs for various models, using
different experimental strategies. The results are encoded in a
detection function $\Sigma$ which indicates to within how many sigmas we
can claim a detection of secondary oscillations, given a particular
model and experiment.

We apply the statistic to both the standard CDM scenario and an open
CDM model which is tuned to confuse inflation and cosmic strings in
all but the existence or otherwise of secondary oscillations.  For a
wide range of experiments we allow the beam size, sky coverage, and
detector noise to vary, and use this framework to compute the
detection function for secondary peaks.

\section{Power spectrum estimation from real observations}

There are several factors affecting how well one can measure the CMBR power
spectrum from real observations, which we now discuss.

(i) Distortion of the underlying spectrum due to the finite size of
the observed field. For (square) fields of size $L \geq 4$ degrees
(suitably windowed with a cosine bell or Hann window) this is not a
severe problem for detecting secondary peak structure.

(ii) Cosmic/sample variance, which places constraints on the
minimum sky-coverage necessary to achieve a given accuracy. Roughly
speaking, if $f_{\rm s}$ is the fraction of sky observed, then
$\sigma^2(C_{\ell})/C_{\ell}^2 \approx 1/(\ell f_{\rm s})$.

(iii) Instrumental noise, which we shall assume is uncorrelated for
simplicity, and characterised by $\sigma_{\rm pix}$, the rms pixel
noise, and $\Omega_{\rm pix}$, the area of a pixel. If we consider the
most general case where only a fraction $f_{\rm s}$ of the sky is
mapped, then for a detector of fixed sensitivity, and for a fixed
total observing time, then by varying $\Omega_{\rm pix}$ and $f_{\rm
s}$ the quantity $w^{-1}=\sigma^2_{\rm pix}\Omega_{\rm pix}
(4\pi/f_{\rm s})$, remains constant, and is therefore an important
qualifier for noise on maps obtained using different scanning
strategies.

(iv) Diffuse foreground emission, which can severely hamper the
measurement of CMBR anisotropies.  A discussion of these foreground
components, and the regions of frequency/multipole space in which each
dominates, is given by \cite{te96}.  The main components of this
foreground are Galactic dust, synchrotron and
free-free emission. Algorithms for separating these components from
the CMBR signal are discussed by \cite{te96} and
\cite{maisinger}. Typically the errors associated with the separation
process are of a similar magnitude to the average errors on an
individual frequency channel due to instrumental noise alone, but
details depend on the separation algorithm used.

(v) Point sources, which cannot be removed from spectral information
alone. This requires the identification of the sources by
higher-resolution observations at a frequency close to that of the
CMBR observations, with sufficient flux sensitivity to indentify all
point sources down to some flux limit roughly equal to the
instrumental noise of the CMBR observations.
We note here that although it is generally
believed that point source contamination becomes less important as the
observing frequency increases above about 100 GHz, there is no direct
evidence for this. Moreover, even the population of radio point
sources at frequencies above about 10 GHz is rather uncertain, and it
may be inadvisable to rely on low frequency surveys such as the 1.5 GHz
VLA FIRST survey \cite{becker} to subtract point sources from
CMBR maps made at much higher frequencies.

\section{Observing Doppler peaks in standard CDM}

The idea is to apply to a particular model a statistic sensitive only
to the existence or absence of secondary oscillations in the power
spectrum. In this section we consider the power spectrum predicted by
the standard inflation/CDM scenario with $\Omega_0=1$, $h_0=0.5$ and
$\Omega_b=0.05$ (which we shall call sCDM).
 
\begin{figure}[t]
\centerline{\epsfig{
file=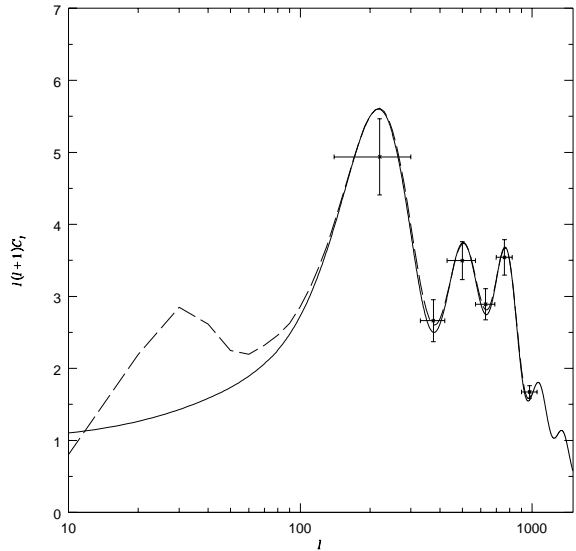,
width=8cm}}
\caption{The ensemble average power spectrum predicted by sCDM 
(solid line), and the ensemble average power spectrum for a square
observing field of size 10 degrees after windowing with a cosine bell
(dashed line). The points indicate 
the corresponding power spectrum estimates for a particular realisation
in each of the bins indicated by the horizontal error bars. The vertical
error bars indicate the theoretical cosmic/sample variance in the absence of
noise.}
\label{mph_fig1}
\end{figure}

To this end we first compute the average broad band power
$C_i$ in each of several equally spaced-bins, denoted by horizontal bars in 
Fig.~\ref{mph_fig1}.
We then infer the convexity ${\cal C}_i$ of the spectrum at each bin position
(apart from the first and last bins) from
${\cal C}_i = (C_{i-1}+C_{i+1})/2-C_i$.
These convexities are all negative if there are no 
secondary peaks, but alternate in sign for sCDM. If the overall error 
in ${\cal C}_i$ is $\sigma^2({\cal C}_i)$ then
one can define an oscillation detection function as 
\begin{equation}
{\Sigma}_i={|{\langle {\cal C}_i\rangle}|\over \sigma
({\cal C}_i)},
\end{equation}
for $i=2$ and $i=4$, which tells us
to within how many sigmas we can claim a detection of secondary peaks.
The method for computing the estimates $C_i$ of the power spectrum in
each bin, and their associated standard errors, taking into account
limited sky-coverage and instrumental noise, are discussed in detail
in \cite{hobmag} and \cite{maghob}.

From Fig.~\ref{mph_fig1} we see that the first dip in the sCDM power
spectrum is more easily detected than than the second one, a situation
only exacerbated by finite resolution and the presence of
instrumental noise.  Therefore we shall confine ourselves to
considering the detection function $\Sigma_2$, which from now on we
refer to simply as $\Sigma$.

\begin{figure}
\centerline{
\epsfig{
file=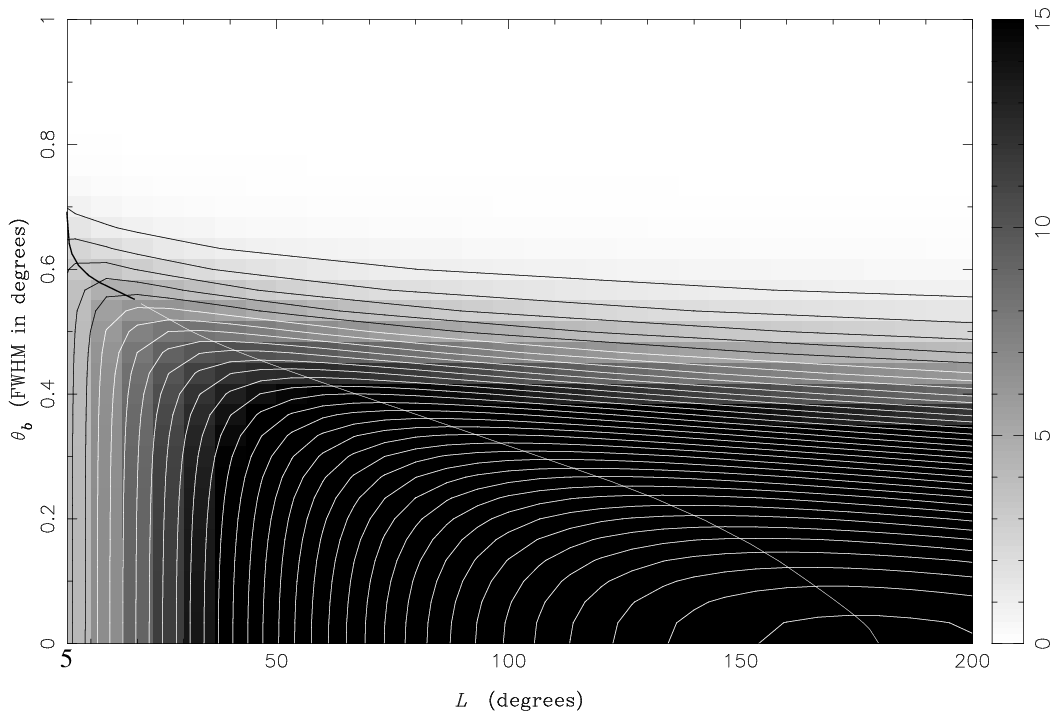,
width=8cm}
\qquad
\epsfig{
file=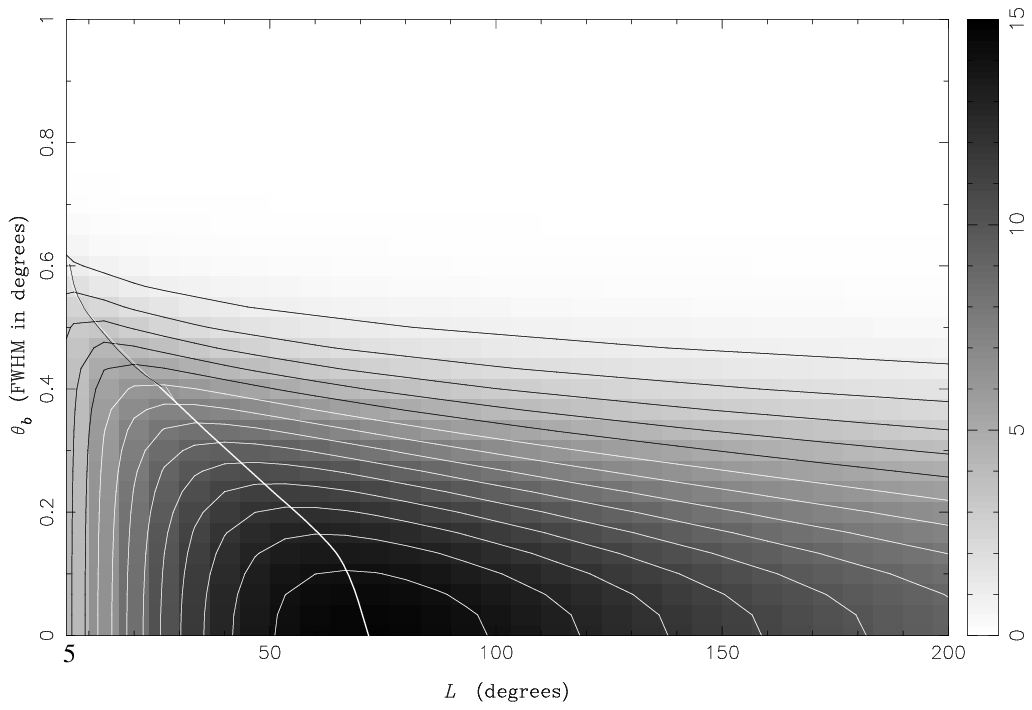,
width=8cm}
}
\caption{(Left) Low noise $w^{-1}=(25\mu K)^{2}(\rm{deg})^{2}$ contours
of the $\Sigma$ function. 
Sky coverage $L$ varies between 5 degrees and all sky ($L=202$ degrees), 
and the beamsize FWHM between 0 and 1 degrees. (Right)
High noise $w^{-1}=(60\mu K)^{2}(\rm{deg})^{2}$
contours of the $\Sigma$ function.}
\label{mph_fig2}
\end{figure}

The detection function $\Sigma=\Sigma(L,\theta_b,w^{-1})$, where $L$
is the linear size of the observed (square) field [the all-sky limit
can be recovered by setting $L^2=4\pi$ sr $\approx $ (202 deg)$^2$],
$\theta_b$ is FWHM of the observing beam, and $w^{-1}$ is the noise
level discussed above. This function is plotted in Fig.~\ref{mph_fig2} for
the low noise case $w^{-1} = (25 \mu\rm{K})^2(\rm{deg})^2$ and
the high-noise case $w^{-1} = (60 \mu\rm{K})^2(\rm{deg})^2$.

For any beamsize there is a maximum sky coverage beyond which the
detection is not improved.  If anything the level of the detection
decreases, but typically not by much.  The ideal scanning strategy is
then defined by a line $L_i(\theta_b)$ which intersects the contours
of $\Sigma$ at the lowest $L$-value at which a plateau has been
achieved in the detection function.  The significance of the detection
obtained for an ideally scanned experiment depends on the beam
size. For example, in the low-noise case,
if $\theta_b=0.6^{\circ}$, the ideal coverage is a
patch of $L_i(0.6^{\circ})=5$ degrees, which results in a 3-sigma
detection.  If $\theta_b=0.5^{\circ}$, on the other hand, an 8-sigma
detection can be obtained with $L_i=35$ degrees.  The detection
provided by an optimally scanned experiment increases at first very
quickly as the beam is reduced below $\theta_b=0.6^{\circ}$ (from
3-sigma at $\theta=0.6^{\circ}$ to 33-sigma at $\theta_b=
0.2^{\circ}$). By reducing $\theta_b$ from $0.2^{\circ}$ to zero,
however, the detection is only increased by 2-sigma (from 33 to 35).
For this level of noise the maximal detection is 35 sigma and is
achieved with $\theta_b<3'$ and all-sky coverage.  
For low noise levels all-sky coverage
is never harmful, but it
is the beamsize that determines how good a detection can be achieved,
and how much sky coverage is actually required for an optimum level of
detection.

For noise levels of the order $w^{-1}=(25\mu K)^{2}(\rm{deg})^{2}$ the
overall picture is always as in Fig.~\ref{mph_fig2}. In particular, there
is always a top contour indicating the maximal detection allowed by the
given noise level. The maximum $\Sigma$ is always achieved with
infinite resolution, but one falls short of this maximum by only a
couple of sigmas if $\theta_b\approx 0.1^{\circ}$.  If the noise is
much smaller than this, however, the summit of $\Sigma$ is beyond
$L=202^{\circ}$. For $w^{-1}=(15\mu
K)^{2}(\rm{deg})^{2}$, for instance, all-sky coverage becomes ideal
for any $\theta_b<0.3^{\circ}$.

If, on the other hand, the noise is much larger than $w^{-1}=(25\mu
K)^{2}(\rm{deg})^{2}$ then the $\Sigma$ contours are qualitatively
different, as shown in Fig.~\ref{mph_fig2} for $w^{-1}=(60\mu
K)^{2}(\rm{deg})^{2}$.
The beamsize is now a crucial factor. A beamsize of
$\theta_b=0.5^{\circ}$ would provide a 3-sigma detection
(with $L_i=10^{\circ}$), but reducing
the beamsize to about $\theta_b=0.4^{\circ}$ improves the detection to
6-sigma (with $L_i=20^{\circ}$). It is also clear from the figure
that, for high noise
levels, forcing all-sky coverage dramatically decreases the detection.

\section{Open CDM models and cosmic strings}

We may repeat the above analysis for different cosmological models.
We therefore consider the case of maximal confusion between
inflation/CDM and cosmic string scenarios by comparing a cosmic
strings model with a CDM model for which the main peak in the power
spectrum has the same position and shape (but the latter exhibits
secondary peaks).  For definiteness we have chosen a CDM theory with a
flat primordial spectrum, $\Omega_0=0.3$, $h_0=0.6$, and $\Omega_b
h_0^2=0.02$.  We shall call this theory stCDM, the CDM competitor of
cosmic strings. As before we simply study the first
dip detection function of stCDM, and then take this detection function
as a cosmic string rejection function.
 
\begin{figure}
\centerline{\epsfig{
file=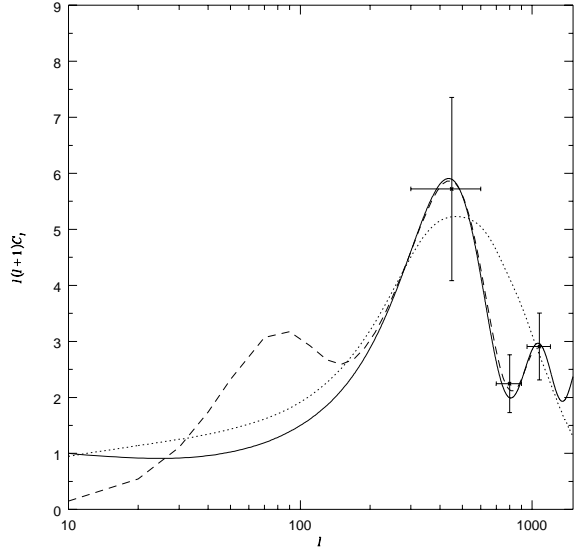,
width=8cm}}
\caption{The angular power spectrum of stCDM (solid line) and one possible
cosmic string scenario (dotted line). The dashed line is the ensemble
average stCDM 
power spectrum as sampled by an experiment with a field size $L=2$.
The points indicate the average power in each bin
for stCDM. The horizontal errorbars denote the width of the bins,
and the vertical errorbars show the sample variance of the power estimates for
such an experiment assuming no instrumental noise.}
\label{mph_fig3}
\end{figure}

\begin{figure}
\centerline{
\epsfig{
file=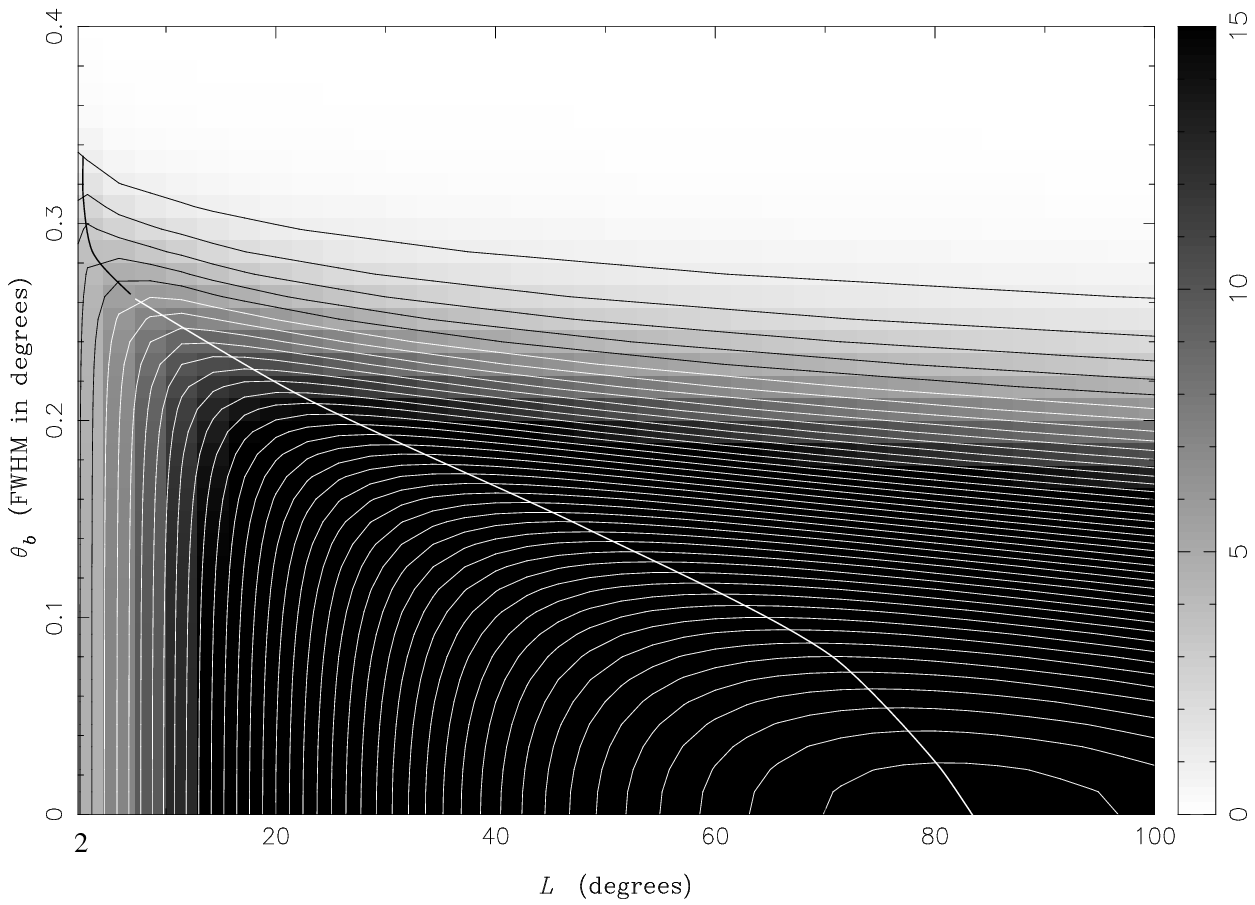,
width=8cm}
\qquad
\epsfig{
file=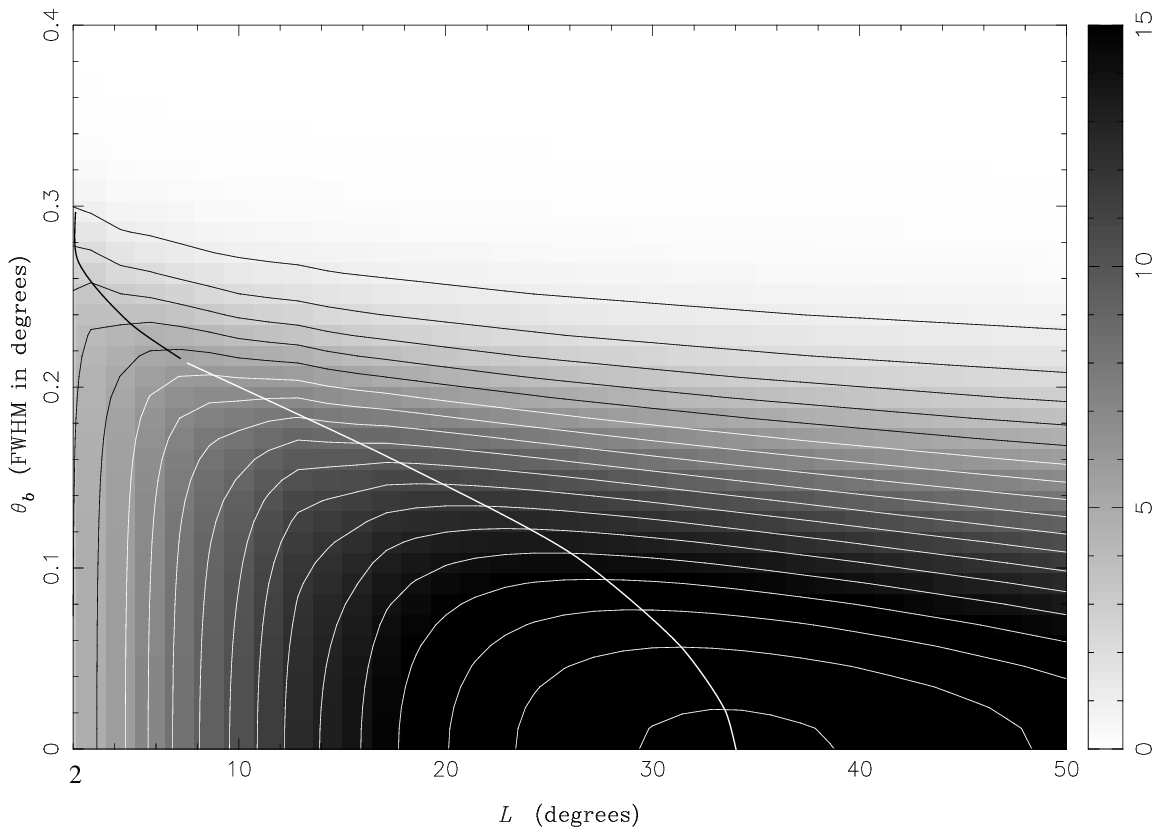,
width=8cm}
}
\caption{(Left) Low noise $w^{-1}=(25\mu K)^{2}(\rm{deg})^{2}$ contours
of the detection function $\Sigma$ for stCDM. (Right) 
High noise $w^{-1}=(60\mu K)^{2}(\rm{deg})^{2}$ contours
of the detection function.}
\label{mph_fig4}
\end{figure}

In Fig.~\ref{mph_fig3} we show the angular power spectrum of stCDM (solid
line) and a possible power spectrum for cosmic strings (dotted line). 
We then simply repeat the same exercise as in the previous section
to obtain the detection function of the first dip of stCDM. The
results are shown in Fig.~\ref{mph_fig4} for the same noise levels as
before. 

Overall we see that in signal dominated regions the detection
is much better for stCDM than for sCDM. This is because features at
higher $\ell$ have a smaller cosmic/sample variance (which
is proportional to $1/\ell$). It can be checked that the cosmic/sample
variance limit, obtained with a single-dish experiment with no noise,
is now $\Sigma\approx 197L^2/(4\pi)$ (as opposed to 
$\Sigma\approx 77L^2/(4\pi)$ for sCDM).
Even in the presence of noise, wherever the signal dominates, the detection
is better for stCDM. However, in noise-dominated regions 
the behaviour of the detection function
for stCDM and CDM
is very different.

The signal-dominated region is greatly reduced in stCDM. Much smaller
beamsizes $\theta_b$ are now required for any meaningful detection. As
shown in Fig.~\ref{mph_fig4}, one would now need
$\theta_b<0.3^{\circ}$ and $\theta_b<0.25^{\circ}$, for noises
$w^{-1}=(25\mu K)^{2}(\rm{deg})^{2}$ and $w^{-1}=(60\mu
K)^{2}(\rm{deg})^{2}$ respectively, in order to obtain a reasonable
detection. Again one can plot an ideal scanning line in the
beam/coverage sections defined by a fixed noise $w^{-1}$. The ideal
sky coverage is much smaller for stCDM than for sCDM. In general the
contours of $\Sigma$ for stCDM compared to sCDM are squashed to lower
$\theta_b$, lower $L$, and achieve higher significance levels, with
steeper slopes. Following an ideal scanning line for any fixed
$w^{-1}$ one reaches a maximal detection allowed by the given level of
noise, which is always better for stCDM than for sCDM. This maximal
detection is normally obtained with a small sky coverage, and infinite
resolution.  Nevertheless, one falls short of this maximum by only a
few sigma if the resolution is about $\theta_b = 0.05^{\circ}-0.1^{\circ}$.  
From Fig.~\ref{mph_fig4}, for $w^{-1}=(25\mu K)^{2}(\rm{deg})^{2}$, one may
now obtain a maximal 43-sigma detection for an ideal scanning area of
$L=65$ degrees.  If $\theta_b=0.1^{\circ}$ a 36-sigma detection is
still obtained.  We also see that a beamsize of $\theta_b <
0.25^{\circ}$ is required to obtain a 3-sigma detection (with $L=4$
degrees), and a 10-sigma detection can be achieved only with
$\theta_b\approx 0.15^{\circ}$ (and $L_i=18$ degrees).  All-sky
coverage for an experiment targeting stCDM is generally inadvisable,
and it would only be optimal for the extremely low level of noise
$w^{-1}<(11\mu K)^{2}(\rm{deg})^{2}$.

\section{Conclusions}

The results obtained here are useful for future CMBR projects in two
different ways.  Firstly they allow the choice of an ideal scanning
strategy (choice of resolution and sky coverage) for detecting
secondary Doppler peaks, given observational constraints such as the
instrumental noise level and the total observing time.  Secondly, one
may compute the expected value of the detection, assuming ideal
scanning, as a function of these parameters.  This provides lower
bounds on experimental conditions for a meaningful detection as well
as an estimate of how fast detections will improve thereafter.

These results also indicate that in order to study Doppler peak
features for sCDM, depending on the noise levels, a large sky coverage
might be desirable, even for a resolution of about
$\theta_b=0.4^{\circ}-0.5^{\circ}$.  If, however, one is instead to test the
high-$l$ opposition between low $\Omega$ CDM and cosmic strings, then
a rather higher resolution is required.  Furthermore, in this context,
all-sky scanning is not only unnecessary, but in fact undesirable.

\acknowledgements{Thanks to Joao Magueijo for many useful discussions.}

\vfill
\end{document}